\documentclass[epj]{svjour}
\usepackage{latexsym}
\usepackage{graphics}
\usepackage{caption}
\usepackage{subcaption}
\usepackage[demo]{graphicx}
\usepackage[title]{appendix}
\usepackage{tabularx}
\usepackage{array}
\usepackage{booktabs}
\usepackage{makecell}

\newcolumntype{Z}{ >{\centering\arraybackslash}X }
\begin{document}\sloppy
\title{Machine learning for buildings characterization and power-law recovery of urban metrics}
\author{Alaa Krayem\inst{1}, Aram Yeretzian\inst{2}, Ghaleb Faour\inst{3} \and Sara Najem\inst{1}% etc
\thanks{\emph{Present address:} sn62@aub.edu.lb}%
}                     % Do not remove
\offprints{}          % Insert a name or remove this line
\institute{Physics Department,  American University of Beirut,
Beirut 1107 2020, Lebanon\and Architecture and Design, American University of Beirut,
Beirut 1107 2020, Lebanon \and National Center for Remote Sensing, CNRS-L, Riad al Soloh, 1107 2260, Beirut, Lebanon
}
\date{Received: date / Revised version: date}
%% The correct dates will be entered by Springer

\abstract{
In this paper we focus on a critical component of the city: its building stock, which holds much of its socio-economic activities. In our case, the lack of a comprehensive database about their features and its limitation to a surveyed subset lead us to adopt data-driven techniques to extend our knowledge to the near-city-scale. 
Neural networks and random forest are applied to identify the buildings' number of floors and construction periods' dependences on a set of shape features: area, perimeter, and height along with the annual electricity consumption, relying a surveyed data in the city of Beirut. The predicted results are then compared with established scaling laws of urban forms, which constitutes a further consistency check and validation of our workflow. %The pipeline we followed consists of a data cleaning the data used to build the algorithms and removing outliers from it. Then, the neural networks are developed for each feature prediction and evaluated with different performance metrics. After validation, the networks are expected to provide reasonable estimates of the missing features for a more complete database of urban buildings. 
\PACS{
      {PACS-key}{discribing text of that key}   \and
      {PACS-key}{discribing text of that key}
     } % end of PACS codes
} %end of abstract
\authorrunning 
 \titlerunning
\maketitle
\section{Introduction}
\label{intro}
A key question for planning, designing, and managing urban spaces is how the different city's components interact and influence its dynamics \cite{McPhearson2016}. %In this context, adopting ecological principles allows for the developing of sustainable processes \cite{Rosenzweig2010}. Urban ecology has emerged as a science of city, that integrates multidisciplinary and transdisciplinary approaches capable of predicting future evolution of urban characteristics to assist the desicion making on how to navigate the transformation of cities towards sustainability \cite{McPhearson2016,Bettignies2019}.
It is therefore important to recognize cities as complex systems with emergent properties, far from equilibrium dynamics, and with energy requirements for self-maintenance \cite{McPhearson2016,Batty2008a}. \\
Urban form is a crucial component of the city which is concerned with its infrastructure's spatial patterns. Its evolution is governed by the rules of competitive processes which manifest themselves in self-similar fractal patterns \cite{Batty2008} or equivalently in scaling laws, which govern the changes of city component with its size \cite{Bettencourt2013,Batty2015,Steadman2009,schlapfer2015urban}. % : a rank-size rule, allometry which is the study of changes in shape with size and scaling 
At the micro-scale, urban buildings, which are described as ``containers of socio-economic activities'' \cite{Ravetz2008} are of particular interest.  A comprehensive building survey data, specifically regarding their uses, ages, and sizes, is essential to support more effective policymaking relating to the sustainable management of cities \cite{Hudson2018}. For instance, much of the work on the building stock has been driven by the energy sector. Particularly, urban buildings accounts for a high portion of the GreenHouse Gas emissions through electricity consumption \cite{Chen2019,Cerezo2013}. Identifying buildings attributes helps in simulating the buildings' energy performance, identify spatiotemporal patterns to assess the impact of retrofitting strategies to reduce energy consumption, and adapt buildings for climate change \cite{Davila2014,Hong2016,Evans2017,Costanzo2019,Krayem2019}. Moreover, tracking the rate of change of cities and the survival of buildings is essential to estimate the distribution and lifecycle of stock material in order to inform on the best practices of its management and utilization or the so-called ''industrial ecology''  \cite{Tanikawa2009,Kohler2009}. \\
\indent Therefore, generating a building database is important for urban science broadly speaking. It basically starts by collecting existing information as in the property taxation database \cite{Bruhns2000} or conducting ground surveys. However, this data is sometimes expensive, unavailable or insufficient. For this reason, taking advantage of new data sources, methods and tools is a central focus area in urban research. Volunteered Geographic Information (VGI) platforms, which are crowdsourcing tools, are gaining emerging interest. Coloring London, where residents are encouraged to fill, substitute, and update the buildings database themselves is one such example \cite{Hudson2019a}. Moreover, the automated capture and extraction of building attribute data is more and more facilitated by the development of computational resources, data-driven learning techniques and remote sensing \cite{Hudson2018,Hu2013,Meinel2009,Belgiu2014}. \\
\indent In this paper, we apply data-driven techniques to assist the collection of a spatial building data. Neural Networks and Random Forests are built to link the physical character of the buildings to their number of floors and vintage respectively. 
We first outline the database we are working on in Section \ref{sec:data}. We then proceed by developing the machine learning algorithms for the buildings' attributes prediction in Section \ref{sec:rd} before illustrating the results in Section \ref{sec:results} and their relation to established urban scaling. 

\section{Data collection and preprocessing}
\label{sec:data}
Beirut is located on the eastern shore of the Mediterranean sea with a stock of $17,742$ buildings. The latter's corresponding footprints' attributes used in this work (area, perimeter and height), were obtained from the National Center for Scientific Research CNRS Lebanon, while additional information on a subset of $7,122$ buildings was surveyed by the Saint-Joseph University. It includes buildings' year of construction, type, number of floors and of apartments. Their corresponding construction years were converted into construction periods based on the city's architectural history, which witnessed five major waves of construction each with specific distinctive features \cite{Arbid2002}. The distribution of the USJ subset according to the year of construction is given in Table \ref{tab:percentage}.
\begin{table}[!htp]\begin{center}
\caption{Percentage of buildings per period of construction in the dataset.}
\label{tab:percentage}
\begin{tabularx}{\columnwidth}{ccc}
 \hline
\hline \noalign{\smallskip}
Construction period&Label&Percentage of buildings\\
 \hline \hline\noalign{\smallskip}
1900-1923& 1 & $1.2\%$\\
1924-1940& 2 & $7.8\%$\\
1941-1960& 3 & $42.1\%$\\
1961-1990& 4 & $39.1\%$\\
After 1991& 5 & $9.7\%$\\
 \hline
  \hline
\end{tabularx}
\end{center}
\end{table}

Moreover, data on the annual electricity consumption for many buildings were obtained from the national power utility: Electricite du Liban (EDL). Entries with missing fields, incorrect buildings' heights $(\leq 2.8m)$, or atypical floor height $(\leq 2.8m$ or  $\geq 4,5m)$ were removed from the dataset. % However, the collected data was prone to inconsistencies such as incomplete and missing fields. Consequently, it was crucial to remove these inconsistencies before the data is used. First, data cleaning was applied by removing buildings with empty entries, height less than $2.8m$ or floor height less than $2.8m$ or greater than $4.5m$ in USJ data. 
The floor height ranges were also determined as in \cite{Arbid2002}.  Residential and mixed buildings types were kept while others such as hospitals, places of worship, and schools were removed, which left us with $1,968$ buildings.  

This step was followed by the application of an Isolation forests (iForest) scheme \cite{Liu2008}, which is an outliers' detection procedure and is essential to the removal of noisy, incorrect or aberrant information in the dataset. The ouliers' removal was based on a set of features: the yearly electricity consumption, floors' number, height, perimeter, area, period of construction and type of the building. Subsequently, $432$ buildings were classified as outliers and were therefore removed from the dataset. The latter then consisted of $1,536$ buildings, which we used in what follows. 
%\begin{figure*}[!htp]
%\centering
%\begin{subfigure}[b]{0.9\columnwidth}
%\includegraphics[width=\columnwidth,keepaspectratio]{rank_size_rule_area.eps}
%\caption{Area}
%\label{fig:ranksize_area}
%\end{subfigure}\hfill
%\begin{subfigure}[b]{0.9\columnwidth}
%\includegraphics[width=\columnwidth,keepaspectratio]{rank_size_rule_height.eps}
%\caption{Height}
%\label{fig:ranksize_floor}
%\end{subfigure}
%\caption{Initial analysis of buildings data.}
%\label{fig:ranksize}
%\end{figure*}

\section{Methods}
\label{sec:rd}
%The algorithms used in this section give rise of configuration of the underlying relationships between the buildings' attributes.

The building' floor number is shown to be dependent on the building's height, area, perimeter, and annual electricity consumption. Having established this dependency, the building's construction period's relation to the aforementioned features is investigated. The selection of the buildings' features, that is the independent variables, can be justified by a correlation analysis achieved with the Pearson coefficient for the floors' number, and a dependency strength achieved with the logistic regresison' accuracy score for the construction period, as seen in Table \ref{tab:pearson}. The accuracy score is given by: $accuracy = ({1}/{n_{samples}})\times{\Sigma_{i=1}^{n_{samples}}1(y_{pred,i}=y_{true,i})}$, where $y_{pred,i}$ is the predicted value of the i-th sample and $y_{true,i}$ is the corresponding true value.

\begin{table}[]
\caption{Pearson coefficient and logistic regression's accuracy score describing, respectively, the correlation and dependency between dependent variables and selected variables for prediction.}
\label{tab:pearson}
\centering
\begin{tabular}{lcc}
 \hline
\hline 
 &  \thead{Pearson\\ Coefficient} &  \thead{Accuracy\\ score} \\ \hline
 &  \thead{Floor\\ number} &  \thead{Construction\\ period} \\  \hline
\hline \noalign{\smallskip}
Electricity consumption & 0.57 & 0.51 \\
Height & 0.95 & 0.59 \\
Area & 0.37 & 0.46 \\
Perimeter & 0.38 & 0.46  \\
 \hline
\hline 
\end{tabular}
\end{table}
For the prediction of number of floors, which is an integer value, a neural network (NN) for a multivalued non-linear regression was trained, whereas for the classification of the construction periods, which is a categorical value with labels ranging from 1 to 5 given in Table \ref{tab:percentage}, NN, random forests, and logistic regression were applied. In order to measure the performance of the NN with different architectures, regression metrics such as the mean absolute error (MAE), the mean squared error (MSE), the mean absolute percentage error (MAPE),  and the coefficient of determination ($R^2$) were computed. Similarly measures of performance of classification algorithms were also computed such as the accuracy score and the f1-score. Further, the resulting models were applied on the test sets to evaluate their performance. Subsequently,  the best performing models were extended to the whole city. %While training the algorithms, it was found that the number of features is limited and is leading to a low accuracy. %In an attempt to improve the performance of the models, 
\subsection{Floors}
The dataset of $1,536$ samples was subdivided into training, validation, and test sets each containing respectively $859$,  $369$, and $308$ samples. The features of the training set were normalized and consequently their values ranged from 0 to 1.  
The NN's hyper-parameter tuning was carried out exploring different numbers of hidden units, neurons, and learning rates. %The results were very similar for the different configurations. 
Additionally, the cumulative distributions of the number of floors $P(f)$ of the 1,536 buildings and that of the combination of the latter set with the predicted 6,877 buildings' floors were computed.  We tested whether these distributions can be explained by power-laws: $P(f) = ({f}/{f_{min}})^{-\alpha+1}$, where $f$ is the floor number, $\alpha$ is the exponent, and $f_{min}$ is the cutoff of the power-law, or whether a lognormal, whose parameters are given in \cite{clauset2009power}, can better explains the distributions. The parameters were determined using the \textit{poweRlaw} package in \textit{R} for a discrete data set and subsequenlty the models were compared using the likelihood ratio test.%Further, in order to get a measure of the uncertainty of the distributions parameters we perform a boostrapping procedure using the bootstrap function. 

\subsection{Period of construction}
Table \ref{tab:percentage} shows that the dataset is highly imbalanced, with only $1.2\%$ of the buildings belonging to the first construction period, compared to $42.1\%$ belonging to the third period. Resampling the data was crucial before proceeding. Since we had a relatively small dataset ($1,536$ samples), oversampling the minority classes of the training set was applied to improve the quality of the predictive model. This was achieved using SVMSMOTE \cite{Nguyen2011} by creating synthetic observations of the minority classes, at each iteration of the cross-validation.
 %Figs. \ref{fig:era_hista} and \ref{fig:era_histb} show the variation of the construction period distribution before and after resampling of the training set.
Different configurations of logistic regression, Random Forest classifiers and NN classifiers were examined and compared with the accuracy score.

%\begin{table*}[!htp]\begin{center}
%\caption{Pearson coefficient describing the correlation between dependent variables and selected variables for prediction.}
%\label{tab:pearson}
%\begin{tabular}{c|*{6}{p{2cm}}}
%\hline\noalign{\smallskip}
%&Floors Number&Construction period&Electricity Consumption&Height&Area&Perimeter\\
%Floors Number& 1 &-&0.57&0.95&0.37&0.38\\{\smallskip}
%Construction period& 0.63 & 1&0.34&0.62&0.22&0.23\\
%%Electricity Consumption&0.34&0.57&1&0.60&0.59&0.58\\
%%Height&0.62&0.95&0.60&1&0.42&0.42\\
%%Area&0.22&0.37&0.59&0.42&1&0.94\\
%%Perimeter&0.23&0.38&0.58&0.42&0.94&1\\
%\end{tabular}
%\end{center}
%\end{table*}

\section{Results}
\label{sec:results}

\subsection{Floors prediction}
\label{sec:floors_prediction}

%The dataset of $1,536$ samples was subdivided into a training set of $859$ samples, a validation set of $369$ samples and a test set of $308$ samples. The training set was normalized between 0 and 1. 
 The input's layer's 4 neurons correspond to the area, perimeter, height, and electricity consumption, while to output layer's single neuron is that of the period of construction. The optimum number of hidden layers and their neurons were found to be 1 and 8 respectively, with a learning rate of $0.01$ and a sigmoid transfer function. The scores of the applied NN on the test are given by: 
\begin{itemize}
\itemsep0em 
\item mean absolute error MAE $= 0.54$
\item mean squared error MSE $= 0.73$
\item mean absolute percentage error MAPE $= 7.2\%$
\item $R^2 = 87.7\%$
\end{itemize}

The prediction of the floors' number for the rest of the city's buildings could now be extended keeping in mind that buildings with missing input features had to be excluded. This left us with $6,877$ buildings whose number of floors is to be predicted. The results along with the surveyed data from USJ were mapped as shown in Fig. \ref{floors_distribution} in Appendix \ref{appendix:appendixA}.

\begin{figure}[!htp]
\begin{center}
\includegraphics[width=0.72\columnwidth, ,keepaspectratio]{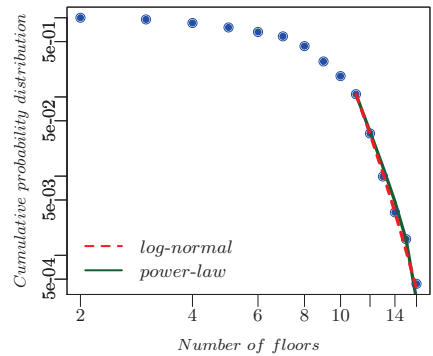}
\caption{$P(f)$ of the 1,536 building buildings is shown blue, the power-law is shown in green, and the lognormal in shown in red. Their respective parameters are  $f_{min}=11$ and $\alpha = 12.92$, while those of the lognormal are given by $f_{min} = 11$, $\mu  = 1.55$, and $\sigma = 0.27$.}
\label{fig:cumul_floors_distribution}
\end{center}
\end{figure}

\begin{figure}[!htp]
\begin{center}
\includegraphics[width=0.72\columnwidth, ,keepaspectratio]{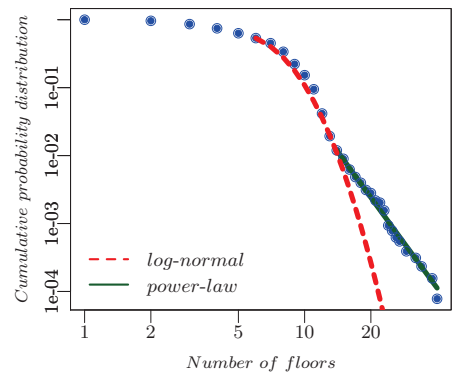}
\caption{$P(f)$ for all the buildings is shown in blue. The green line is the power-law with  $f_{min}=14$, and $\alpha = 5.35$. while the red line corresponds to the lognormal with $f_{min} = 6$ params $= 2.05 ,0.28$. }
\label{fig:cumul_floors_distribution_all}
\end{center}
\end{figure}

The cumulative distribution of the number of floors of the USJ buildings was evaluated. Additionally, the latter along with predicted buildings' floor number was also computed. They are shown respectively in Figs. \ref{fig:cumul_floors_distribution} and \ref{fig:cumul_floors_distribution_all}. 
In the first, the ratio $r$ of the log-likelihoods of the data between the power-law and lognormal is negative, which means that the lognormal is a better fit, while in the second $r>0$ indicating that the power law with exponent $\alpha = 5.35$ is a better fit. This latter parameter is in accordance with the findings of \cite{Batty2008b}, where the exponent of the height distribution of London was shown to be $\alpha = 5.26$.

\subsection{Period of construction prediction}
\label{sec:era}
\begin{figure}[!htp]
\begin{center}
\includegraphics[width=1.1\columnwidth, ,keepaspectratio]{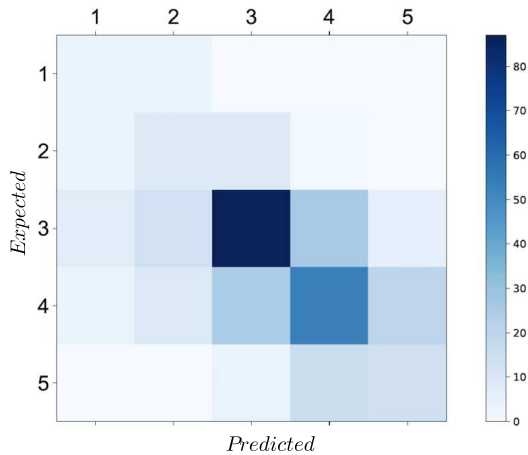}
\caption{Distribution of buildings per predicted period of construction in Beirut administrative area.}
\label{fig:confusion}
\end{center}
\end{figure}

The exhaustive hyper-parameter tuning and models cross-comparison converged to a random forest with 100 decision trees with an accuracy score of $56.7\%$. Further, its accuracy score on the test set was given by $48.7\%$. Using this model, despite its low accuracy, the rest of Beirut's buildings were tagged with their corresponding predicted construction period (Fig. \ref{era_distribution} in Appendix \ref{appendix:appendixA}). 
Further, the confusion matrix was plotted in Fig. \ref{fig:confusion}. It revealed that the algorithm best predicted the first construction period with an accuracy of $62.5\%$, while its worse accuracy was attained with the second construction period with only $35.9\%$.

%\begin{figure}[!htp]
%\begin{center}
%\includegraphics[width=0.9\columnwidth, ,keepaspectratio]{era_hist.eps}
%\caption{Data distribution before oversampling.}
%\label{fig:era_hista}
%\end{center}
%\end{figure}

%\begin{figure*}[!htp]
%\centering
%\begin{subfigure}[b]{0.9\columnwidth}
%\includegraphics[width=\columnwidth,keepaspectratio]{era_hist.eps}
%\caption{Before oversampling}
%\label{fig:era_hista}
%\end{subfigure}\hfill
%\begin{subfigure}[b]{0.9\columnwidth}
%\includegraphics[width=\columnwidth,keepaspectratio]{era_hist_sampled.eps}
%\caption{After oversampling}
%\label{fig:era_histb}
%\end{subfigure}
%\caption{Construction periods distribution in the training set of the USJ data before and after oversampling to effectively handling imbalanced datasets.}
%\label{fig:era_hist}
%\end{figure*}

\section{Discussion}
\label{sec:discussion}
In the previous sections, we have presented a pipeline which relies on machine learning to complement urban buildings' database. 
It should be noted that our starting point was a data whose floors distribution is best described by a lognormal however the combination of the latter with the predicted data was shown to be a power-law, which is in full accordance with the measured one for other cities \cite{Batty2008b}. This is further a consistency check on the validity of the results. 
%The developed models can have important implications in urban ecology studies.
High accuracy in predicting buildings' floors number was attained revealing a strong relation with the building height, area, perimeter and electricity consumption. The quantification of the floors is relevant to energy planning, as it helps simulating the energy demand by representing each floor as a thermal zone. A floor can be further subdivided into subzones for more accuracy of the building performance simulation \cite{Dogan2017c}. Furthermore, it can help approximating the building population for micro-scale modeling and analysis of human behavior \cite{Greger2015}.\\

\indent On the other hand, the period of construction could be predicted with an accuracy score of $48.7\%$ only.  More training data may be required. However, the low accuracy may be related to the need for more variables on which the period of construction depends, such as window to wall ratio (WWR), wall thickness and other era-specific descriptors. The construction period gives insight into the materials of the buildings, which can inform materials flows and stocks models for valuation of buildings, as well as the determination of their energy performance and refurbishment techniques \cite{Aksoezen2015,Ng2014}, and the identification of future waste streams along with recovery strategies \cite{Heinrich2019}.

\section{Conclusion}
\label{sec:summary}
Finally, we developed NN algorithms to predict the number of floors and the construction period of buildings given their heights, areas, perimeters and electricity consumption. We begun by cleaning the available dataset and removing unreliable entries and outliers. Then, we evaluated the significance of each input feature on the output to justify its selection. The NN was able to predict the number of floors with a high prediction accuracy with a coefficient of determination of $R^2$ of $87.7\%$. Then, the construction period's Random Forest was built after re-sampling of the data to overcome its imbalance. Subsequently, the exponent of the power-law governing the floor distribution was shown to be conforming with that appearing in the literature.    

\section{Authors contributions}
%All the authors were involved in the preparation of the manuscript.
%All the authors have read and approved the final manuscript.
A. Krayem performed data preprocessing and cleaning, A. Yeretzian specified the architectural construction waves, S.Najem, A. Yeretzian, and G. Faour designed the workflow, S. Najem and A. Krayem wrote the machine algorithms, and all authors contributed to the writing of the manuscript. 
\begin{appendices}
\section{}\label{appendix:appendixA}
\begin{figure*}[!htp]
\begin{center}
\includegraphics[width=1.1\columnwidth, ,keepaspectratio]{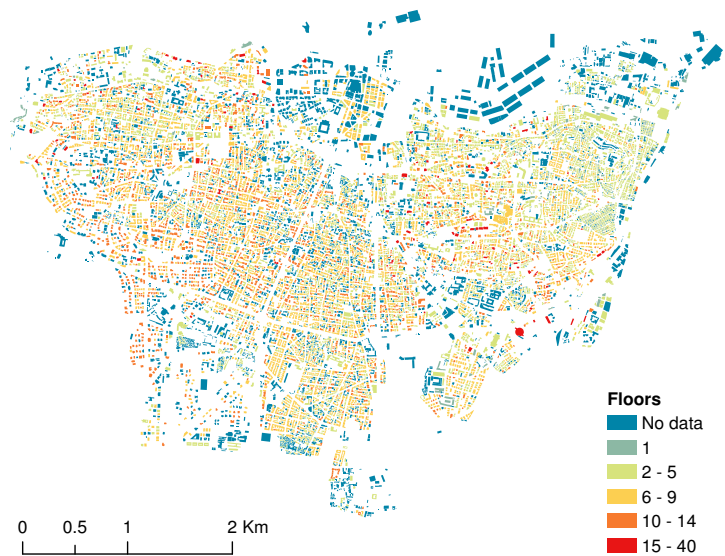}
\caption{Distribution of buildings per floors' number in Beirut administrative area.}
\label{floors_distribution}
\end{center}
\end{figure*}

\begin{figure*}[!htp]
\begin{center}
\includegraphics[width=1.1\columnwidth, ,keepaspectratio]{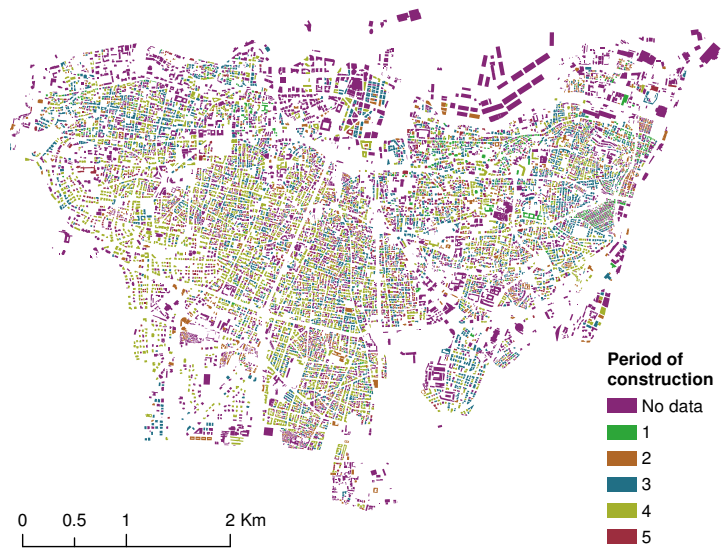}
\caption{Spatial distribution of buildings per period of construction in Beirut administrative area.}
\label{era_distribution}
\end{center}
\end{figure*}

\end{appendices}
%\section{Acknowledgments}

%%
%% The section below may be edited at your convenience to acknowledge 
%% each author's contribution to the manuscript.
%% You may remove it if you are a single author.
%%

%https://tex.stackexchange.com/questions/434142/bib-file-how-to-change-the-position-of-the-year-field-in-the-references-epj-bs
 %BibTeX users please use
 \newpage
\bibliographystyle{epj}
\bibliography{library}

%% Non-BibTeX users please use
%\begin{thebibliography}{}
%
% %and use \bibitem to create references.
%Author, \textit{Book title} (Publisher, place year) page numbers
%\end{thebibliography}

\end{document}